\documentclass[amsmath,prl,aps,twocolumn]{revtex4}
\usepackage{amsthm,amsfonts,graphicx,verbatim}

\newcommand{\Tr}{{\rm Tr}}
\newcommand{\tr}{{\rm tr}}

\def\req#1{(\ref{#1})}
\newcommand{\D}{{\rm d}}

\begin{document}
\title{Lower entropy bounds and particle number fluctuations in a Fermi sea}
\author{Israel Klich}
\email{klich@tx.technion.ac.il} \affiliation{Department of
Physics, Technion - Israel Institute of Technology, Haifa 32000
Israel}
\date{June 2003}
\begin{abstract}
In this Letter we demonstrate, in an elementary manner, that given
a partition of the single particle Hilbert space into orthogonal
subspaces, a Fermi sea may be factored into pairs of entangled
modes, similar to a BCS state. We derive expressions for the
entropy and for the particle number fluctuations of a subspace of
a Fermi sea, at zero and finite temperatures, and relate these by
a lower bound on the entropy. As an application we investigate
analytically and numerically these quantities for electrons in the
lowest Landau level of a quantum Hall sample.
\end{abstract}
 \maketitle

The study of quantum many particle states, when measurements are
only applied to a given subsystem, are at the heart of many
questions in physics. Examples where the entropy of such
subsystems is interesting range from the quantum mechanical
origins of black hole entropy, where the existence of an event
horizon thermalizes the field density matrix inside the black hole
\cite{Bombelli,CallanWilczek}, to entanglement structure of spin
systems \cite{Vidal}. In this work we address the relation of
entanglement entropy of fermions with the fluctuations in the
number of fermions.

First we show that given a subspace of the single particle Hilbert
space, a Fermi sea may be factorized into pairs of entangled modes
in and out of the subspace, thereby writing the state as a BCS
state \footnote{A treatment of a BCS factorization for Gaussian
states was carried out by A. Botero and B. Reznik (preprint
quant-ph/0404176)}.

While upper bounds on entropy, were a subject of numerous
investigations, especially since Bekenstein's
bound\cite{BekensteinBound}, lower bounds on entropy are less
known.

We show that given a "Fermi sea", the entropy of the ground state,
restricted to a particular subspace $A$ of the single particle
space, relates to the particle number fluctuations in the subspace
via the inequality:
\begin{eqnarray}\label{EIneq}
S_A\geq 4\log2 \Delta N^2_A\geq -8\log 2 \ll N^4_A\gg
\end{eqnarray}
Where $S_A$ is the entropy associated with the subspace $A$, and
$\Delta N^2_A$ are the fluctuations in the particle number in $A$,
and $\ll N^4_A\gg$ is the fourth cummulant of particle number
\footnote{Note that a fourth cummulant may have either sign}. The
importance of this result lies in the fact that particle
fluctuations are, in principle, easier to measure, and are
fundamentally related to the quantum noise in various systems. On
the technical side note that the right-hand side, has the
advantage of being easy to calculate analytically in a wider class
of problems.

We start by examining the ground state of non-interacting
fermions, in arbitrary external potential, when measurements are
applied to a given part of the space. The basic example is a Fermi
sea or a Dirac sea where we are interested in the relative entropy
of a given region of space, and in fluctuations in the number of
particles there, but one may also consider entanglement in Fermion
traps (Fermi degeneracy of potassium atoms ($^{40}K$) has been
observed by De Marco and Jin\cite{MarcoJin}).

The discussion is also relevant for systems which behave like a
non-interacting Fermi gas, as in problems of transport at the zero
temperature limit\footnote{The zero temperature is completely
degenerate. Otherwise, we need the condition (for strongly
degeneracy) $T\ll E_F$}, in ideal metals, where transport may be
approximated well within a non-interacting theory, due to good
screening. In an ideal single channel conductor the analogy is
done by mapping excitations that travel at the Fermi velocity to a
time - energy coordinate representation in discussion of quantum
pumps \cite{MartinLandauer,LevitovLeeLesovik,AvronTimeEnergy}. The
analogy is especially manifest in the problem of switching noise
\cite{KlichLevitov}.

The ground state of a noninteracting Fermi gas, containing $N$
particles is obtained by occupying the allowed states $\phi_i\in
H$ ($H$ is the single particle Hilbert space) up to energy $E_f$,
i.e.:
\begin{eqnarray}
|gs>=\prod_{E(\phi_i)<E_f} \psi^{\dag}(\phi_i)|0>
\end{eqnarray}
Where $\psi^{\dag}$ are creation operators which satisfy the usual
canonical anti-commutation relations (CAR):
\begin{eqnarray}
&
[\psi(\phi_i),\psi^{\dag}(\phi_j)]_{+}=<\phi_i,\phi_j> \\
\nonumber &
[\psi(\phi_i),\psi(\phi_j)]_{+}=[\psi(\phi_i)^{\dag},\psi^{\dag}(\phi_j)]_{+}=0
\end{eqnarray}

The state $|gs>$ or the "Fermi sea", is a typical ground state for
a large class of Hamiltonians (sometimes it is necessary to carry
a suitable Bogolubov transformation). It can also describe spin
chains via the Jordan-Wigner transformation.

Let $A$ be a subspace of the single particle Hilbert space $H$, so
that $H=A\oplus A^{\perp}$ and let $E=span\{\phi_i;1\leq i\leq
N\}$ be the subspace of occupied single particle states of $H$
(the Fermi sea).

Let $P_A$ be the orthogonal projection on $A$.  Consider the
matrix $M(A)_{ij}=<P_A\phi_j,P_A \phi_i>$,
$i,j=1,..,N$\footnote{We restrict ourself to states that obey
$|P_A \phi_j>|^2>0$, otherwise they are already of the required
form.}. $M_{ij}$ is a hermitian matrix so that it can be
diagonalized by a unitary $U$: $M=U^{\dag}{\rm diag}(d_i)U$.

The new orthonormal modes are defined by
$$A_l={\sum_k U^{\dag}_{lk}P_A\phi_k\over \sqrt{d_l}}$$ where
the factor $d_l$ is the l-th eigenvalue of $M$, and serves to
normalize the $A_l$ with the inner product on $H$. Similarly we
take $B_l={\sum_k U^{\dag}_{lk}P_A^{\perp}\phi_k\over
\sqrt{1-d_l}},$ which are orthonormal since $M(A^{\perp})=I-M(A)$.
Obviously $A_i\in A$ and $B_i\in A^{\perp}$.

Since $U$ is unitary we write $E=span\{U^{\dag}\phi_i\}$. Using $
U^{\dag}_{il}\phi_l=\sqrt{d_i}A_i+\sqrt{1-d_i}B_i$, the ground
state may be written as:
\begin{eqnarray}\label{factorization} &
|gs>=\prod_{i=1}^{N} \Psi^{\dag}(U^{\dag}_{ij}\phi_j)|0>=\\
\nonumber & \prod_{i=1}^{N}
(\sqrt{d_i}\Psi^{\dag}(A_i)+\sqrt{1-d_i}\Psi^{\dag}(B_i))|0>
\end{eqnarray}
Note that the $0\leq d_i\leq 1$ due to the structure of $M$. In
the particular case that $[P_E,P_A]=0$ then the exercise is
trivial, as they can be diagonalized simultaneously. In addition,
whenever there is a symmetry operator that commutes with $P_A$ and
$P_E$, then the resultant eigenmodes are invariant under the
symmetry.

The ground state may be written also as a BCS state in the
following way. Consider the vector $|A>=\prod
\psi^{\dag}(A_i)|0>$, defined by "filling" the modes in $A$, and
redefine $\psi(A_i)=\psi^{\dag}_h(A_i)$, then:
\begin{eqnarray}
|gs>=\prod_i(\sqrt{d_i}-\sqrt{1-d_i}\psi^{\dag}(B_i)\psi^{\dag}_h(A_i))|A>
\end{eqnarray}
Similar to a BCS state, where the entangelment structure of the
modes pairs $A_i,B_i$ is clearly seen.

If we factor the single particle Hilbert space $H$ into
\begin{eqnarray}
H=\bigoplus_{i=1}^{N}span\{A_i,B_i\}\bigoplus(Complement),
\end{eqnarray}
the fermion Fock space factors into an appropriate tensor product.
We can write naturally the density matrix as:
\begin{eqnarray}
\rho=|gs><gs|=\bigotimes_{i=1}^{N}|\nu_i><\nu_i|,
\end{eqnarray}
where
$\nu_i=(\sqrt{d_i}\Psi^{\dag}(A_i)+\sqrt{1-d_i}\Psi^{\dag}(B_i))|0_i>$.
After a partial trace over the $B's$, the reduced density matrix
is given by:
$\rho_A=\bigotimes_{i=1}^{N}\left(%
\begin{array}{cc}
  1-d_i & 0 \\
  0 & d_i \\
\end{array}%
\right)$,
where $d_i$ is the probability of having a particle in mode $A_i$.

From the density matrix $\rho_A$ we have:
\begin{eqnarray}\label{2cum}&
<N>_A=\Tr(M),
\\ \nonumber &
\Delta N^2_A=<N^2-<N>^2>_A=\Tr M(1-M)\, .
\end{eqnarray}
To study further moments, denote $P(k)$ the probability of having
$k$ fermions in $A$ and consider the generating function:
\begin{eqnarray}
\chi(\lambda)=\sum P(k)e^{i\lambda k}=\det(1+M(e^{i\lambda}-1)).
\end{eqnarray}
The cummulants of the number of fermions in $A$ may be extracted
from the generating function by differentiating $\log\chi$ and
setting $\lambda=0$. For example, the fourth cummulant is given
by:
\begin{eqnarray}\label{4cum}&
\ll N^4\gg=\partial^{4}_{i\lambda}\log\chi|_{\lambda=0}=\\
\nonumber & \Tr(M(1-M)(1-6M+6M^2))
\end{eqnarray}

The reduced entropy is obtained from $\rho_A$:
\begin{eqnarray}\label{entropy}
S_A= -\Tr( M\log M+(1-M)\log(1-M)).
\end{eqnarray}
This entropy is equivalent to the usual entropy of a Fermi gas
with occupation number operator $M$.

\begin{figure}
\includegraphics[scale=0.4]{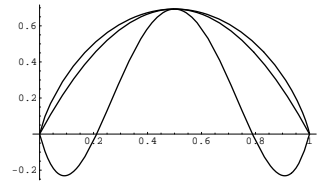}\caption{The functions
$ -x\log x-(1-x)\log(1-x)\geq 4\log 2x(1-x)\geq -8\log 2
x(x-1)(1-6x+6x^2)$ are related to the entropy, 2nd and 4th
cummulants}\label{cummulants}
\end{figure}

Comparing the expressions \req{2cum},\req{4cum} and \req{entropy},
and using the fact that $ 0\leq M\leq 1$, we get the inequalities
\req{EIneq} as promised (Fig. \ref{cummulants}).

The difference between the cummulants and the entropy comes from
eigenstates with probability away from $1/2$ but not exactly $0$
or $1$. The fourth cummulant may be used to estimate the
contribution to the second moments of eigenvalues of $n_A$ away
from $1/2$, as the function $-2x(x-1)(1-6x+6x^2)$ is negative
outside the interval $(1/2-\sqrt{3/4},1/2+\sqrt{3/4})$

The inequalities are valid for finite temperatures too. To see
this, consider a general density matrix of the form
$\rho=Z^{-1}e^{-K_{ij}a^{\dag}_ia_j}$. By tracing out the
$A^{\perp}$ degrees of freedom, the reduced density matrix
acquires the form
\begin{eqnarray}
\rho_A=\det({1-n_A})e^{\log({n_A\over 1-n_A} )_{ij}a^{\dag}_ia_j}.
\end{eqnarray}
Where $n_A=P_A{1\over 1+e^K}P_A$ is as an operator on $A$. The
construction of this distribution from the covariance matrix
$<a_ia^{\dag}_j>$ was discussed in several papers, mainly in the
context of density matrix renormalization group
\cite{Peschel,CheongHenley}, where a similar expression was
obtained.

The resulting density matrix on $A$ may be considered as thermal
by appropriately choosing the "energies"
$\epsilon_i=\log{1-d_i\over d_i}$. The density matrix may then be
written as a "thermal" state $\rho={Z_A^{-1}}e^{-\sum \epsilon_i
\psi^{\dag}(A_i)\psi(A_i)}$. Note that while the new eigenstates,
may have large values for the energy with respect to $K$, the
operator $n_A$ is well defined, and bounded by 1. The entropy is:
$$S=-\Tr (n_A\log n_A+(1-n_A)\log(1-n_A))$$
The generating function $\chi(\lambda)$ becomes:
\begin{eqnarray}
\chi(\lambda)=\det(1+n_A(e^{i\lambda}-1)),
\end{eqnarray}
and the expressions for the moments are the same with the
identification: $M\rightarrow n_A$.

One may attempt to find a similar inequality for Bosons. For a
closed system in thermal equilibrium
\begin{eqnarray}
S=\Tr ((1+n_{BE})\log(1+n_{BE})-n_{BE}\log n_{BE})
\end{eqnarray}
(where $n_{BE}$ is a Bose-Einstein occupation operator) and
\begin{eqnarray}
(\Delta N)^2=\Tr n_{BE}(1+n_{BE}).
\end{eqnarray}
In most cases particle fluctuations will dominate the entropy,
however one can check that whenever $n_{BE}<1$ (a dilute boson
gas) then:
\begin{eqnarray}\label{BosINEQ}
S_{Bosons}\geq (\log 2)(\Delta N)^2_{Bosons}
\end{eqnarray}
For systems in which the single particle energies have a gap
$\Delta$ above the zero energy, as the temperature goes to zero we
have $n_i\rightarrow 0$, making the inequality above valid.

While the cases of spin chains and free fields, were studied in
numerous works, other important examples wait addressing. Here we
consider the problem of 2D electrons in a quantum Hall sample.

The filled lowest Landau level states are spanned by
$|k>={1\over\sqrt{\pi k!}}z^k e^{-|z|^2/2}$\footnote{In the
symmetric gauge for the vector potential $A_i = -{B\over 2}
\epsilon_{ij}x^j$. The length unit is $\sqrt{2}l$ where
$l=\sqrt{\hbar c\over eB}$ is the magnetic length.}.

In \cite{Balachadran}, the entanglement entropy of a Chern-Simons
theory describing a quantum hall defined on a disk was shown to
scale like the radius. Let us now calculate the particle
fluctuations. We choose $A$ to be a disc of radius $R$. Due to the
radial symmetry of the system and of $A$, the factorization
\req{factorization} is possible mode by mode, and the lowest
Landau level may be written as:
\begin{eqnarray}
|LLL>=\prod_{k}(\sqrt{d_k}\psi^{\dag}(|k>_A)+\sqrt{1-d_k}\psi(|k>_{A_{\perp}}))|0>
\end{eqnarray}
Where $d_k=1-{\Gamma(1+k,R^2)\over k!}$ are given in terms of the
incomplete gamma function, $|k>_A={1\over
\sqrt{d_k}}\chi(|z|<R)|k>$, and $|k>_{A_{\perp}}$ is defined
similarly \footnote{Note that the average number of particles
contained in $A$ is $\sum d_k\sim R^2$.}.
\begin{figure}
\includegraphics*[scale=0.5]{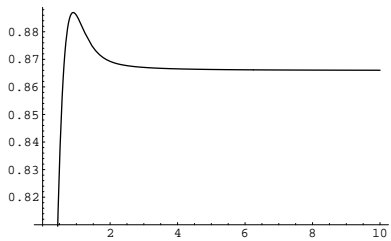}\caption{$S(R)/\Delta N(R)^2$ for a disc in the lowest landau level}\label{HallFluc}
\end{figure}
In this case that both the entropy and particle fluctuations are
asymptotically linear in $R$, thereby proportional to the boundary
area.

Particle fluctuations in this model are given by \\ $<\Delta
N_{A}^2>=\sum_l M_{ll}(1-M_{ll})$. This sum may be approximated
analytically as follows: write
\begin{eqnarray}&
\sum_l M_{kk}^2=\sum_{k,l}\delta_{k,l}M_{k,l}=
\\ \nonumber &  {2\over
\pi}\int_0^{2\pi}\D\theta\int_0^{R}\int_0^{R}\D x\D y e^{x^2
e^{-i\theta}+y^2 e^{i\theta}} x y e^{-x^2-y^2}
\end{eqnarray}
Which can be written, after the $x,y$ integrations, as the contour
integral:
$$
\sum_l M_{kk}^2=
 -{1\over
2\pi i}\oint_{|z|=1}\D z
{(1-e^{-R^2(1-z)})(1-e^{-R^2(1-1/z)})\over (1-z)^2}
$$
Since the integrand is analytic outside $z=0$ we break it up and
do part of the integration on a contour with
$|z|\rightarrow\infty$:
\begin{eqnarray*}&
(\Delta N(R))^2=
 R^2-{1\over
2\pi i}\oint_{|z|\rightarrow\infty}\D z {e^{-R^2}(e^{R^2 z}-e^{R^2
1/z})\over (1-z)^2}\\ \nonumber & -{1\over 2\pi i}\oint_{|z|=1}\D
z {1-e^{-R^2 (2-z-1/z)})\over (1-z)^2}
\end{eqnarray*}
The first integral can be evaluated in the limit
$R\rightarrow\infty$, by the residue theorem, and simply cancels
the $R^2$ term. The contribution to the remaining integral comes
from around the point $\theta=0$, we linearize near this point and
get, as $R\rightarrow\infty$:
\begin{eqnarray}&
(\Delta N(R))^2\sim {1\over 2\pi }\int_{-\infty}^{\infty}\D \theta
{1-e^{-R^2 \theta^2}\over \theta^2}= {1\over \sqrt{\pi} }R
\end{eqnarray}
We see that in this model, the fluctuations are proportional to
the boundary. A numerical check shows that the entropy and
$<\Delta N^2_A>$ are comparable (Fig. \ref{HallFluc}).

\begin{figure}
\includegraphics*[scale=0.5]{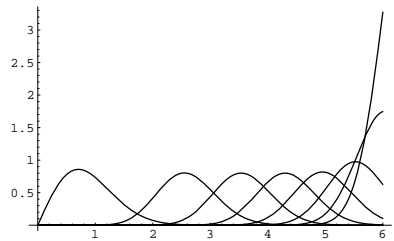}
\caption{The modes $k=6n$, $n=0,..7$ for $R=6$, are concentrated
closer to the boundary with increasing $k$} \label{HallFlucModes}
\end{figure}

In this example it is manifest that for a fixed $R$, as $k$ grows
larger (especially\footnote{ In this problem there is a crossover
near $k=R^2$, where $d_k\sim{1/2}$. To see this use the property
${\Gamma(n,n)\over\Gamma(n)}<1/2<{\Gamma(n,n-1)\over\Gamma(n)}$,
from which $d_{R^2}(R^2)<1/2<d_{R^2}(R^2-1)$ (valid for $R^2$
integer)}. for $k>R^2$), the modes $|k>_A$ are localized stronger
at the boundary, showing that the information is concentrated near
the boundary (Fig. \ref{HallFlucModes}). Note that in this case we
consider from the outset an infinite number of fermion modes, in
contrast with the spin-chain case, where a finite segment yields a
finite dimensional fermionic Fock space after the Jordan-Wigner
transformation.

We remark that since quasi-free fermionic states are related to
determinantal processes \cite{Soshnikov}, equivalent inequalities
may be obtained for the connection of the entropy and the
fluctuations in such systems. In particular, if one is interested
in particle fluctuations, then the theorem of Lebowitz and Costin
\cite{CostinLebowitz}, ensures a gaussian behavior for the scaled
particle number fluctuations as the volume grows large for a large
class of determinantal processes.

In the case of a 1D Fermi gas, both entropy and particle
fluctuations in a box of size $L$ scale as $\log{k_F L}$
\footnote{when disorder is present we expect the particle
fluctuations to behave like $S=\log l_{e} k_F$, where $l_e$ is the
elastic mean free path.} Particle number fluctuations in 1D
conductors may be observed using ultra-fast transistors, however
it will not be possible to ignore the Coulomb interaction. We note
that for a 1D conductor, there is a pre-factor due to the cross
section, which means that the fluctuations may be quite large.

In this Letter we have shown how the ground state of fermions can
be factored into sets of pairs of modes - inside and outside a
given subsystem. We have outlined the connection between the
entanglement entropy and the fluctuations in the particle number
in two ways: First, the inequality \req{EIneq} supplies a lower
bound on the available entanglement entropy, and second, noting
that in some cases (namely the 1d Fermi sea and the lowest
Landau level) both quantities scale in a similar way.\\
 {\bf Acknowledgment:} I thank J. D. Bekenstein, J. Lebowitz, L. S. Levitov,  B. Shapiro, A.
Retzker and M. Reznikov for useful remarks.


\begin{thebibliography}{99}
\bibitem{Bombelli}
L. Bombelli, R.K. Koul, J. Lee, and R. D. Sorkin, Phys. Rev. D34
(1986) 373.

\bibitem{CallanWilczek}
C.G. Callan and F. Wilczek, Phys. Lett. B 333, 55 (1994).

\bibitem{Vidal}
G. Vidal, J. I. Latorre, E. Rico and A. Kitaev, Phys. Rev. Lett.
90, 227902 (2003)

\bibitem{BekensteinBound}
J. D. Bekenstein, Phys. Rev. D 23, 287 (1981).

\bibitem{MarcoJin}
B. DeMarco and D. S. Jin, Science 285 (1999) 1703.

\bibitem{MartinLandauer}
Th. Martin and R. Landauer, Phys. Rev. B45, 1742 (1992);

\bibitem{LevitovLeeLesovik}
L. S. Levitov, H.W. Lee, and G.B. Lesovik, J. Math. Phys. 37, 4845
(1996).

\bibitem{AvronTimeEnergy}
 J.E. Avron, A. Elgart, G.M. Graf and L. Sadun, J. Math. Phys. 43, 3415-3424 (2002).

\bibitem{KlichLevitov}
I. Klich and L. Levitov, in preparation.

\bibitem{Balachadran}
A.P. Balachandran, L. Chandar, A. Momen , Int. J. Mod. Phys. A12
(1997) 625-642.

\bibitem{Soshnikov}
A. B. Soshnikov Journal of Statistical Physics, Vol. 100, Nos.
3/4, 2000;
R. Lyons, Pub. Mathématiques. de L'IHÉS, Volume 98, Number 1
(2003), 167. - 212.

\bibitem{CostinLebowitz}
O. Costin and J. L. Lebowitz,  Phys. Rev. Lett. 75(1):69-72
(1995).

\bibitem{Peschel}
I. Peschel, J.Phys.A: Math. Gen. 36, L205 (2003);
M. C. Chung, I. Peschel, Phys. Rev. B 64, 064412 (2001);

\bibitem{CheongHenley}
S. Cheong, C. L. Henley, preprint cond-mat/0307172.

\end{thebibliography}
\end{document}